\documentclass[aps,prc,preprint,floatfix,showpacs,amsmath,amssymb,superscriptaddress]{revtex4}
\newcommand{\sectionreferences}{}
\newcommand{\footnotespecial}[2][1]{\footnotetext[#1]{#2}}
\newcommand{\br}{}
  \newcommand{\co}{(Color online)}
\newenvironment{tabular-rdddcdc}{\begin{tabular}{r|dd|dc|dc}}{\end{tabular}}
\newenvironment{tabular-rdddcdcc}{\begin{tabular}{r|dd|dc|dc|c}}{\end{tabular}}
\newenvironment{tabular-cdddd}{\begin{tabular}{c | d d | d d}}{\end{tabular}}
\newenvironment{tabular-cdrdr}{\begin{tabular}{c|dr|dr}}{\end{tabular}}
\usepackage{graphicx}
\usepackage{dcolumn}
\usepackage{color} 
  \def\nuc#1#2{\relax\ifmmode{}^{#1}{\protect\text{#2}}\else${}^{#1}$#2\fi}
  \def\itnuc#1#2{\setbox\@tempboxa=\hbox{\scriptsize\it #1}
    \def\@tempa{{}^{\box\@tempboxa}\!\protect\text{\it #2}}\relax
    \ifmmode \@tempa \else $\@tempa$\fi}
  
  \newcommand{\beq}{\begin{equation}}
  \newcommand{\eeq}{\end{equation}}
  \newcommand{\bea}{\begin{eqnarray}}
  \newcommand{\eea}{\end{eqnarray}}

  \newcommand{\cte}{\nuc{10}{C}}
  \newcommand{\ctw}{\nuc{12}{C}}
  \newcommand{\cfo}{\nuc{14}{C}}
  \newcommand{\csi}{\nuc{16}{C}}
  \newcommand{\cei}{\nuc{18}{C}}
  \newcommand{\ctwe}{\nuc{20}{C}}
  
  \newcommand{\nm}{\ensuremath{N_\mathrm{max}}}
  \newcommand{\ho}{\ensuremath{\hbar \Omega}}
  
  \newcommand{\nn}{\ensuremath{N\!N}}
  \newcommand{\nnn}{\ensuremath{N\!N\!N}}
  
 \newcommand{\cdb}{CDB2k}

 \newcommand{\nlonn}{chiral $N\!N$}
 \newcommand{\nlonnn}{chiral $N\!N\!+\!N\!N\!N$}
 \newcommand{\inoy}{INOY}
 \newcommand{\tp}{\ensuremath{2^+}}
 \newcommand{\betz}{\ensuremath{B\left(\mathrm{E2;}\;
       2^+_1 \rightarrow 0^+_1\right)}}
\begin{document}

\title{Systematics of $2^+$ states in {C}
  isotopes from the no-core shell model}
\author{C. Forss\'en} 
\email[]{christian.forssen@chalmers.se}
\affiliation{Fundamental Physics, Chalmers University of Technology, 412
  96 G\"oteborg, Sweden} 
\author{R. Roth} 
\affiliation{Institut f\"ur Kernphysik, Technische Universit\"at
  Darmstadt, 64289 Darmstadt, Germany}%
\author{P. Navr\'atil} 
\affiliation{TRIUMF, 4004 Wesbrook Mall, Vancouver, British Columbia,
  V6T 2A3 Canada\footnote{Present address.}\\ 
Lawrence Livermore National Laboratory, P.O. Box 808, L-414, Livermore,
CA 94551, USA} 
\date{\today}
\begin{abstract}
  We study low-lying states of even carbon isotopes in the range
  $A=10-20$ within the large-scale no-core shell model (NCSM). Using
  several accurate nucleon-nucleon (\nn) as well as \nn\ plus
  three-nucleon (\nnn) interactions, we calculate excitation energies of
  the lowest $2^+$ state, the electromagnetic $B(\mathrm{E2;} \;
  2^+_1\rightarrow 0^+_1$) transition rates, the $2^+_1$ quadrupole
  moments as well as selected electromagnetic transitions among other
  states. Recent experimental campaigns to measure $2^+$-state lifetimes
  indicate an interesting evolution of nuclear structure that pose a
  challenge to reproduce theoretically from first principles. Our
  calculations do not include any effective charges or other fitting
  parameters. However, calculated results extrapolated to infinite model
  spaces are also presented. The model-dependence of those results is
  discussed. Overall, we find a good agreement with the experimentally
  observed trends, although our extrapolated $B(\mathrm{E2;} \;
  2^+_1\rightarrow 0^+_1$) value for \nuc{16}{C} is lower compared to
  the most recent measurements. Relative transition strengths from
  higher excited states are investigated and the influence of \nnn\
  forces is discussed. In particular for \nuc{16}{C} we find a
  remarkable sensitivity of the transition rates from higher excited
  states to the details of the nuclear interactions.
\end{abstract}
\maketitle
%
\section{\label{sec:intro}Introduction}
Electric quadrupole (E2) matrix elements are important quantities in
probing nuclear structure. In particular, they are very sensitive to
nuclear deformation, the decoupling of proton and neutron degrees of
freedom, and they are often affected by small components of the nuclear
wave functions. In this paper we perform systematic studies of
observables obtained from diagonal and non-diagonal E2 matrix elements
for even carbon isotopes, from \cte\ to the very neutron rich
\ctwe. Quadrupole moments, corresponding to diagonal E2 matrix elements,
are inherently difficult to measure for excited \tp\
states. Off-diagonal matrix elements, however, have recently been
studied for several unstable carbon isotopes using lifetime
measurements~\cite{McCutchan:2012in,Imai:2004p99,Wiedeking:2008p100,Ong:2008p96,
  Elekes:2009p82,Petri:2011p230}. In this way, the reduced transition
probability, \betz, can be extracted since it's inversely proportional
to the lifetime of the \tp\ state. As a result of these experimental
studies, different claims have been made on the nuclear structure in
this chain of isotopes. Initial excitement was triggered by the
observation of a strongly quenched E2 transition in
\csi~\cite{Imai:2004p99}. Based on the liquid-drop model, which predicts
the $B$(E2) to be inversely proportional to the \tp\ excitation energy,
Imai \emph{et al.}~\cite{Imai:2004p99} claimed an anomalous reduction of
the E2 strength when comparing \tp\ lifetimes for \cfo\ ($E_{\tp} =
7.01$~MeV) and \csi\ ($E_{\tp} = 1.77$~MeV). However, the \csi(\tp)
lifetime was remeasured by Wiedeking \emph{et
  al.}~\cite{Wiedeking:2008p100} providing a much shorter value, thus
indicating a larger $B$(E2) strength. Their results were analyzed in
terms of shell-model calculations. Adjusting the effective neutron
charge to reproduce their measured lifetimes they made the claim that
the results for \csi\ are ``normal'' to this region. Lifetime
measurements of \nuc{16,18}{C} were reported by Ong \emph{et
  al.}~\cite{Ong:2008p96}. The presented results for \csi\ came from a
reanalysis of the original data~\cite{Imai:2004p99}, now giving a larger
but still quenched $B$(E2) strength, while the new \cei\ data indicated
the persistence of the quenching of E2 strengths in heavy carbon
isotopes. Possible explanations were put forward in terms of the
decoupling of protons and neutrons resulting in very low values for the
neutron effective charges and/or the appearance of a new proton magic
number $Z=6$ in this region.  Some of these statements were backed up by
new shell-model calculations by Fujii~\emph{et
  al.}~\cite{Fujii:2007p171} reproducing the \nuc{16,18}{C} results
employing exceptionally small effective charges. An alternative
explanation in terms of core polarization effects was recently proposed
by Ma~\emph{et al.}~\cite{Ma:2010p253}. They used a microscopic
particle-vibration approach to compute core polarization effects on
valence nucleons. In contrast with empirical effective charges, usually
employed in shell-model calculations, they noted a very strong quenching
from core polarization on $sd$-shell neutrons for heavy carbon isotopes.

These various developments provide a strong motivation to perform
large-scale calculations, with realistic interactions, to study the
evolving nuclear structure in the carbon chain of (even) isotopes with
particular focus on \tp\ states and quadrupole moments. We have,
therefore, carried out no-core shell model
(NCSM)~\cite{Navratil:2000p42,Navratil:2009p24} calculations for
low-lying states of the even-even carbon isotopes with $A=10-20$. As
described in more detail below, these
calculations are performed starting from realistic Hamiltonians without
adjustable parameters. In particular, since our many-body scheme does
not involve an inert-core approximation we use bare charges when
evaluating electromagnetic observables.
\subsection{\label{sec:theor}Theoretical formalism}
The NCSM method has been described in great detail in
several papers, see e.g., Refs.~\cite{Navratil:2009p24,Barrett:2013-69}. Here,
we just outline the approach as it is applied in the present study.
We start from the intrinsic Hamiltonian for the $A$-nucleon system $H_A
= \mathcal{T}_\mathrm{rel} + \mathcal{V}$, where
$\mathcal{T}_\mathrm{rel}$ is the relative kinetic energy and
$\mathcal{V}$ is the sum of nuclear and Coulomb interactions. The
potential term will always contain two-body operators, but we can also
include three-body terms originating from an initial \nnn\ force, or
three-body terms induced by a unitary transformation of the
Hamiltonian. This transformation, further described below, is employed
to soften the Hamiltonian for use in a truncated many-body basis.

In this work we have used several different nuclear potentials. Common
to all of them is that they reproduce \nn\ phase shifts with very high
precision.  First we have two pure \nn\ interactions: CD-Bonn
2000~\cite{Machleidt:2001p192} (\cdb), based on one-boson exchange
theory, and INOY~\cite{Doleschall:2004p193} that introduces a
nonlocality to include some effects of three-nucleon forces. The latter
is fitted also to three-nucleon observables.  In addition, we have used
the most recent chiral \nn\ plus \nnn\ interaction, i.e. the N$^3$LO
\nn\ interaction of Ref.~\cite{Entem:2003p194} and a local chiral
N$^2$LO \nnn\ potential with low-energy constants determined entirely in
the three-nucleon system~\cite{Gazit:2009p107}. The regulator cutoff
energy of these chiral potentials is 500~MeV.

We solve the many-body problem in a large but finite harmonic-oscillator
(HO) basis truncated by a maximal total HO energy of the $A$-nucleon
system. The many-body model space is usually characterized by the
truncation parameter \nm, giving the maximum number of HO excitations above
the unperturbed $A$-nucleon ground state. The diagonalization of the
Hamiltonian in this many-body basis is a highly non-trivial problem
because of the very large dimensions that is encountered.  To solve this
problem, we have used a specialized version of the shell model code
\textsc{Antoine}~\cite{Caurier:1999p73}, adapted to the
NCSM~\cite{Caurier:2001p196}. For the runs involving explicit \nnn\
interactions we used the NCSD code~\cite{Navratil:2011unpub} as well as
the \textsc{NSuite} package \cite{Roth:2009p210,Roth:2011p211}, which is
also capable of performing the importance-truncated NCSM calculations
described below.

Due to the strong short-range correlations generated by the \nn\
potentials, we usually compute an effective interaction to speed up the
convergence. Two different similarity transformations have been used to
construct the effective interactions: For \cdb\ and INOY as initial \nn\
interactions we compute two-body effective interactions appropriate to
the low-energy basis truncation by a unitary transformation in the
two-nucleon HO basis (Okubo-Lee-Suzuki effective
interaction~\cite{Navratil:2000p42,Suzuki:1980p56,Okubo:1954-12}). We note that the
approximation of performing the transformation in two-body space, hence
neglecting effective many-body terms, will actually disappear in the
infinite model-space limit.  For the \nlonnn\ Hamiltonian we employ the
similarity renormalization group (SRG) with the initial and induced
three-body terms included
consistently~\cite{Roth:2011p211,Jurgenson:2009p229}. Induced four-body
terms are neglected, and have actually been shown to give non-negligible
contributions to ground-state energies in heavy $p$-shell
nuclei~\cite{Roth:2011p211}. Note also that we will not apply the
unitary transformation to other operators than the Hamiltonian. In
particular, results from long-range operators such as E2 are not
expected to be much affected by this
transformation~\cite{Stetcu:2005p254,Anderson:2010p131}.

Dealing with systems having up to 20 nucleons it is a challenging
task. To push beyond the full \nm-space limit we employ the
importance-truncated (IT) NCSM
scheme~\cite{Roth:2009p210,Roth:2007p116}. It makes use of the fact that
many of the basis states are irrelevant for the description of a set of
low-lying states. Based on many-body perturbation theory, one can define
a measure for the importance of individual basis states and discard
states with an importance measure below a threshold value, thus reducing
the dimension of the matrix eigenvalue problem.  Through a sequence of
IT calculations for different thresholds and an a posteriori
extrapolation of all observables to vanishing threshold, we can recover
the full NCSM results up to extrapolation errors \cite{Roth:2009p210}.
%
\section{\label{sec:resul}Results}
\subsection{\label{sec:energy}Convergence and finite model-space results}
The largest model spaces that we are able to reach in
the full \nm-space NCSM calculations span from $\nm=10$ in \cte, via $\nm=8$
in \nuc{12,14}{C}, to $\nm=6$ in \nuc{16,18}{C} and $\nm=4$ in
\ctwe. The largest matrix dimension was $D=1.4 \times 10^9$ for
\cei. However, using the IT-NCSM scheme we are able to obtain results
also with $\nm = 8$ for \nuc{16,18}{C} and $\nm = 6$ for \ctwe. 

\begin{figure}[hbt]
  \centering
  \includegraphics*[width=15cm]
      {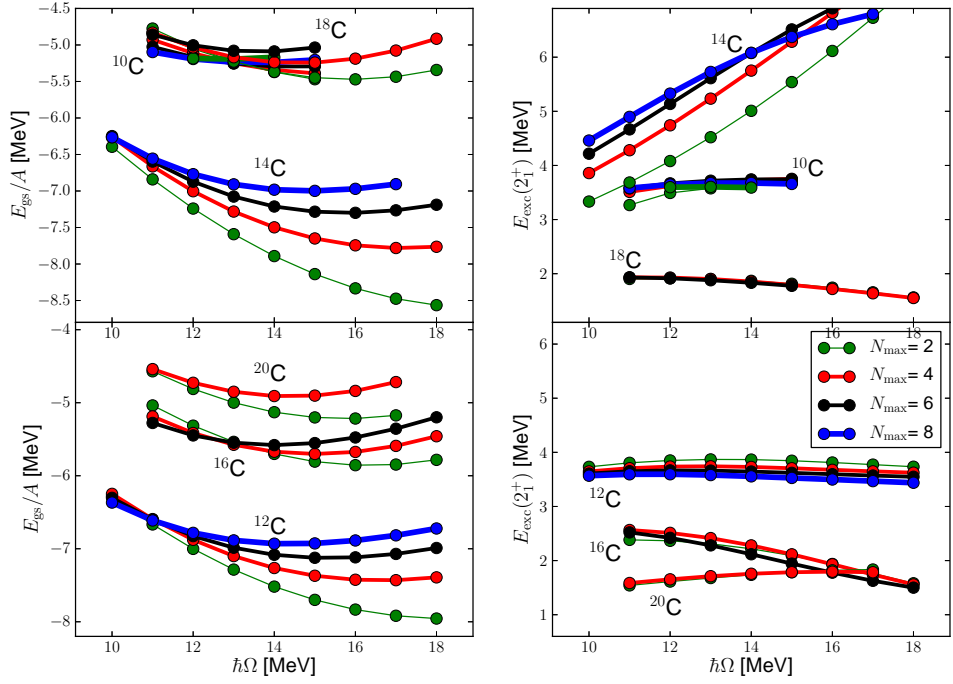}
      \caption{\co\ $\hbar\Omega$-dependence for the ground-state energy
        (presented as $E_\mathrm{gs}/A$)
        and the first \tp\ excitation energy for \nuc{10-20}{C}. Results
        are obtained using the \cdb\ interaction and each
        curve corresponds to a particular model space represented by the
        truncation parameter \nm.%
  \label{fig:z6-hw}}
\end{figure}

Our results exhibit dependence on \nm\ and \ho\ that should disappear
once a complete convergence is reached.  For our detailed studies of
observables we are looking for the regions in which the \nm-convergence
is the fastest and the dependence on \ho\ is the smallest. This optimal
frequency range can vary between different observables and different
isotopes. We will use the \nm-dependence of the binding energy in the
largest model spaces as our primary criterion for selecting the optimal
frequency range. From plots such as the left panels of
Fig.~\ref{fig:z6-hw}, focusing in particular on the trend for large
model spaces, we find that the \ho-range 10-14~MeV is optimal for all
considered observables using the \cdb-interaction and for the whole
range of carbon isotope.

\begin{figure}[thb]
  \centering
  \includegraphics*[width=8cm]
      {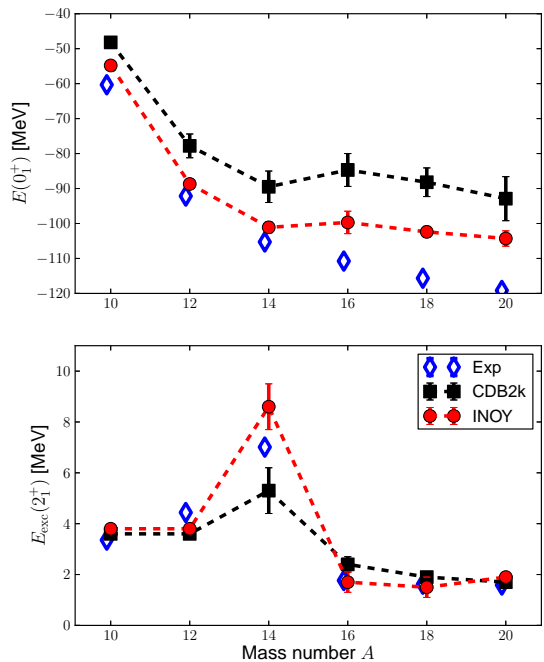}
\caption{\co\ NCSM calculated binding energies and \tp\ excitation
  energies of \nuc{10-20}{C} compared with experimental
  results. %
  \label{fig:cAenergies}}
\end{figure}

We note that relative energies, such as the excitation energy of the
\tp\ state, are well converged, with the possible exception of the
\nuc{14}{C} \tp\ energy. The extraction of energies is not the prime
concern of this paper. In any case, it is clear that the \cdb\ potential
will underbind these isotopes. Still, we can hope that energy
difference, i.e. excitation energies, are relatively well reproduced.
As a rough estimate we can estimate the uncertainty $\Delta E$ of our
results by observing the rates of convergence with respect to
model-space size and HO frequency, respectively.  Such error bars are
obtained using the scheme of Ref.~\cite{Forssen:2009p48}, and shown in a
comparison of experimental data with NCSM calculated binding energies and \tp\
excitation energies of \nuc{10-20}{C} in Fig.~\ref{fig:cAenergies}. It is clear that the
\cdb\ interaction underbinds these isotopes by 10-20~\% while the \inoy\
interaction provides additional binding. The positive two-neutron
separation energy for \csi\ is not reproduced with any of these two
realistic \nn\ interactions. However, the many-body HO basis still
provides a bound-state approximation to these states, and the additional
binding provided, e.g., by \nnn\ interactions will not necessarily
change their structure (see the discussion in
Sec.~\ref{sec:spectrum}). We note that excitation energies are well
converged, and we find a very good agreement with the experimental
trend. The possible exception is the large $2^+_1$ excitation energy of
\cfo\ that is over-predicted with the \inoy\ interaction.

\begin{figure}[hbt]
  \centering
  \includegraphics*[width=15cm]
      {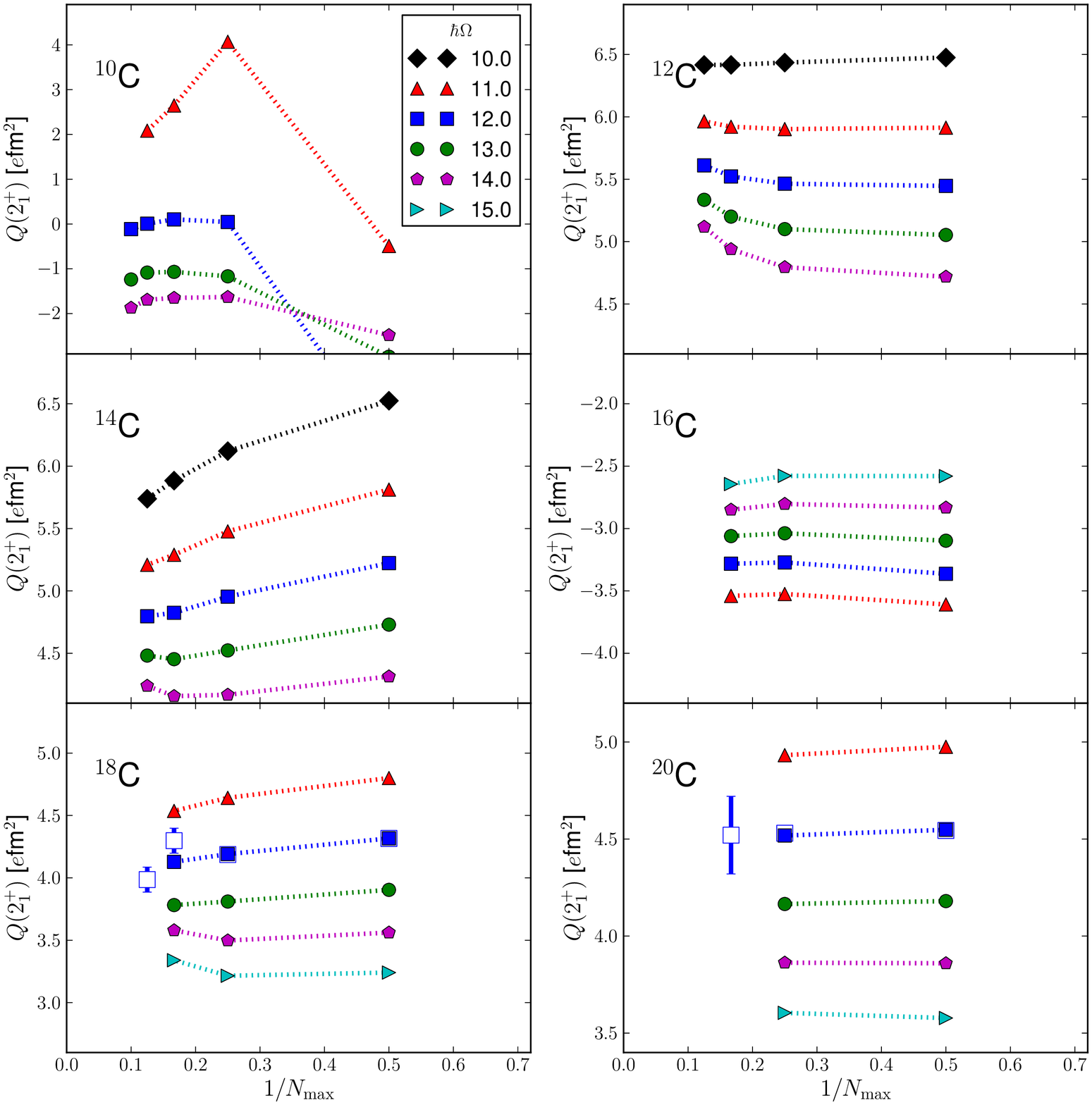}
\caption{\co\ NCSM calculated electric quadrupole moments of the first
  \tp\ states in \nuc{10-20}{C}. Results obtained with
    the \cdb\ \nn\ potential are presented. Filled (open) symbols correspond to full
    (importance-truncated) space results.  See also Table~\ref{tab:qbe2}. %
    \label{fig:cAq}}
\end{figure}

\begin{figure}[hbt]
  \centering
  \includegraphics*[width=15cm]
      {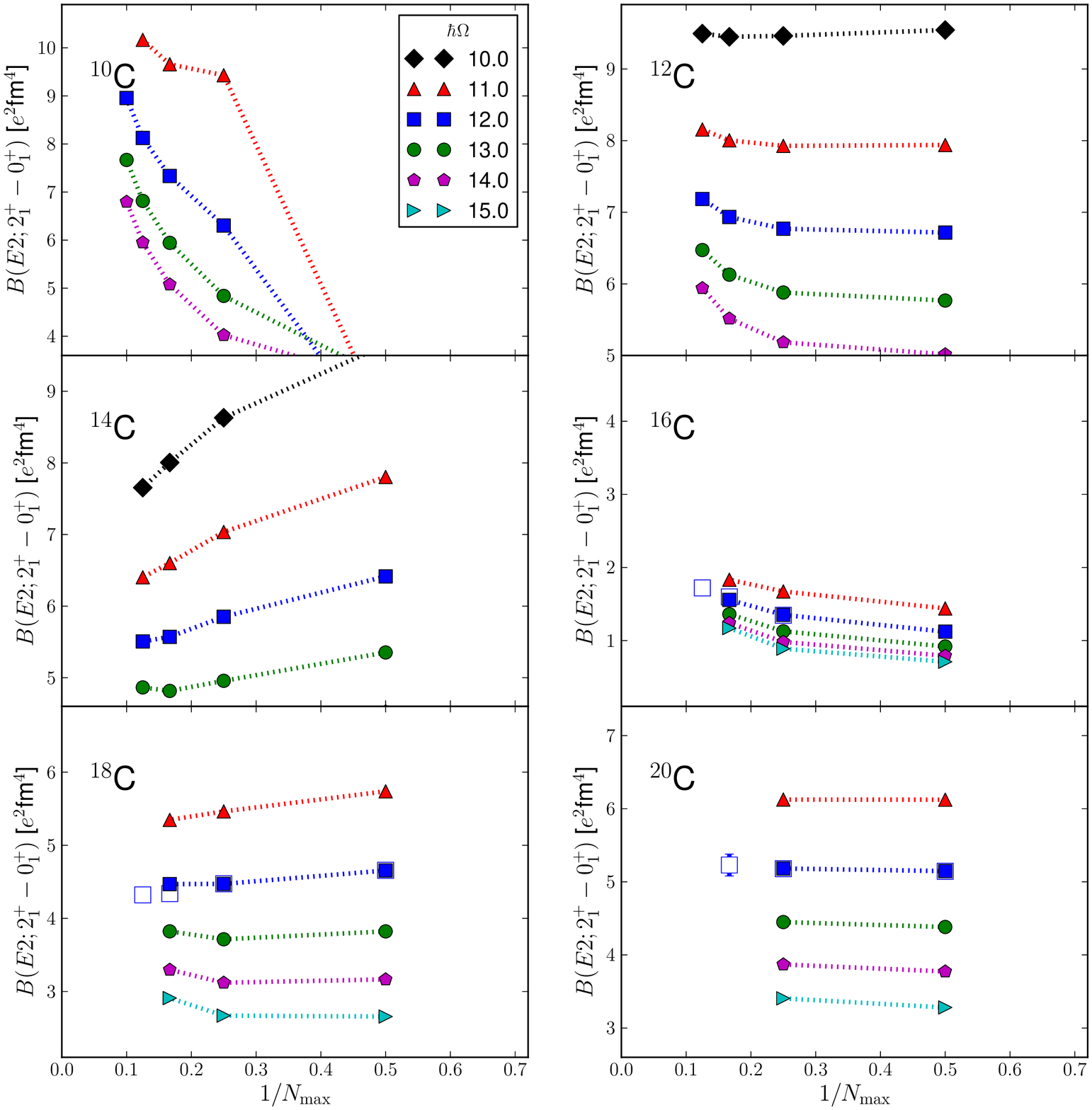}
\caption{\co\ NCSM calculated \betz\ strengths in
  \nuc{10-20}{C}. Results obtained with the \cdb\ \nn\ potential are
  presented. Filled (open) symbols correspond to full
  (importance-truncated) space results.  See also
  Table~\ref{tab:qbe2}. %
  \label{fig:cAbe2}}
\end{figure}

We focus next on electric quadrupole moments, both diagonal and
off-diagonal (transition strengths). Our full \nm-space results are
shown in Figs.~\ref{fig:cAq},\ref{fig:cAbe2} (filled symbols) together
with IT-NCSM results (open symbols).  The data is plotted as a
function of $1/\nm$ for the selected range of HO frequencies. Infinite
model space corresponds to $1/\nm \to 0$.

A reliable extrapolation to infinite model space would be still more
useful. Extrapolation schemes, in particular for quadrupole observables,
are introduced and discussed in more detail in the next section.
\subsection{\label{sec:extrapolation}Extrapolation to infinite model space}
Working in a large many-body HO basis facilitates the implementation of
relevant symmetries and enables us to capture the important physics of
the realistic nuclear interactions that are employed in our \emph{ab
  initio} approach to nuclear structure. However, the need to truncate
our model space introduces constraints on our ability to describe very
short-distance (high energy) as well as long-distance correlations. This
fact was recently expressed in terms of various definitions of infrared
(IR) and ultraviolet (UV) cutoffs in finite HO
bases~\cite{Coon:2012-86, Furnstahl:2012eu}. Suggestions on how to
employ these model-space parameters to extrapolate to infinite model
space were proposed.

Common to all extrapolation schemes is the need to introduce a number of
fit parameters to extrapolate the \emph{ab initio} results to the
infinite space. Using the insights offered by introducing UV and IR
regulators rather then model-space parameters \nm\ and \ho, the number
of free parameters can be significantly reduced. In general,
extrapolations based on pure phenomenology have more fit
parameters~\cite{Maris:2009kp, Forssen:2008p49,Forssen:2009p48}. Dealing
with Okubo-Lee-Suzuki transformed results, for which the Hamiltonian is
\emph{model-space dependent} and the variational principle does not
apply, our experience is that multiparameter fits seem to be necessary.

We focus here on our results for electric quadrupole observables. In
Figs.~\ref{fig:cAq},\ref{fig:cAbe2} we show our full
(importance-truncated) \nm-space NCSM results as filled (open) symbols
for various HO frequencies.  As observed, they exhibit dependence on
\nm\ and \ho. We know that, by construction, this dependence should
disappear once a complete convergence is reached. This implies that
\nm-sequences obtained at different HO frequencies should all converge
to the same result. This feature can be utilized to perform a
constrained fit to multiple
sequences~\cite{Forssen:2008p49,Forssen:2009p48}. To this end, we use as
large an \nm\ basis as feasible for a wide range of HO frequencies, and
extrapolate calculated observables to infinite space.
Results obtained for a range of frequencies are used in the fits. We
find that the convergence behavior for E2 observables, as a function
of \nm, can be rather well fitted by: $q = q_\infty + c_0 / \nm + c_1 /
\nm^2$. The parameters $c_0$ and $c_1$ are allowed to vary for each
\ho-sequence, while the single parameter $q_\infty$ gives the
extrapolated result at $\nm \to \infty$. Typically we use a range of
five HO frequencies for these constrained fits. Finally, an error
estimate is made based on repeating the constrained fit keeping various
subsets (pairs and triples) of frequencies in the selected range.

In addition, we perform an additional, simplified, extrapolation
procedure using simple first-degree polynomials in $1/\nm$: $q = c_0 +
c_1 / \nm$. This allows simple fits to pairs of data $\left( q_{\ho,\nm},
  q_{\ho,\nm-2} \right)$. Using a sequence of such fits (for different
values of \ho) gives a range of $c_0$ parameters that together provides
an estimate for the range of the desired observable $q_\infty$. As we
follow the convergence with increasing model space sizes: $\nm = (8,6), \,
(6,4), \, (4,2)$, we expect to see that this range gets smaller and
smaller. 

\begin{figure}[tb]
  \centering
  \includegraphics*[width=15cm]
      {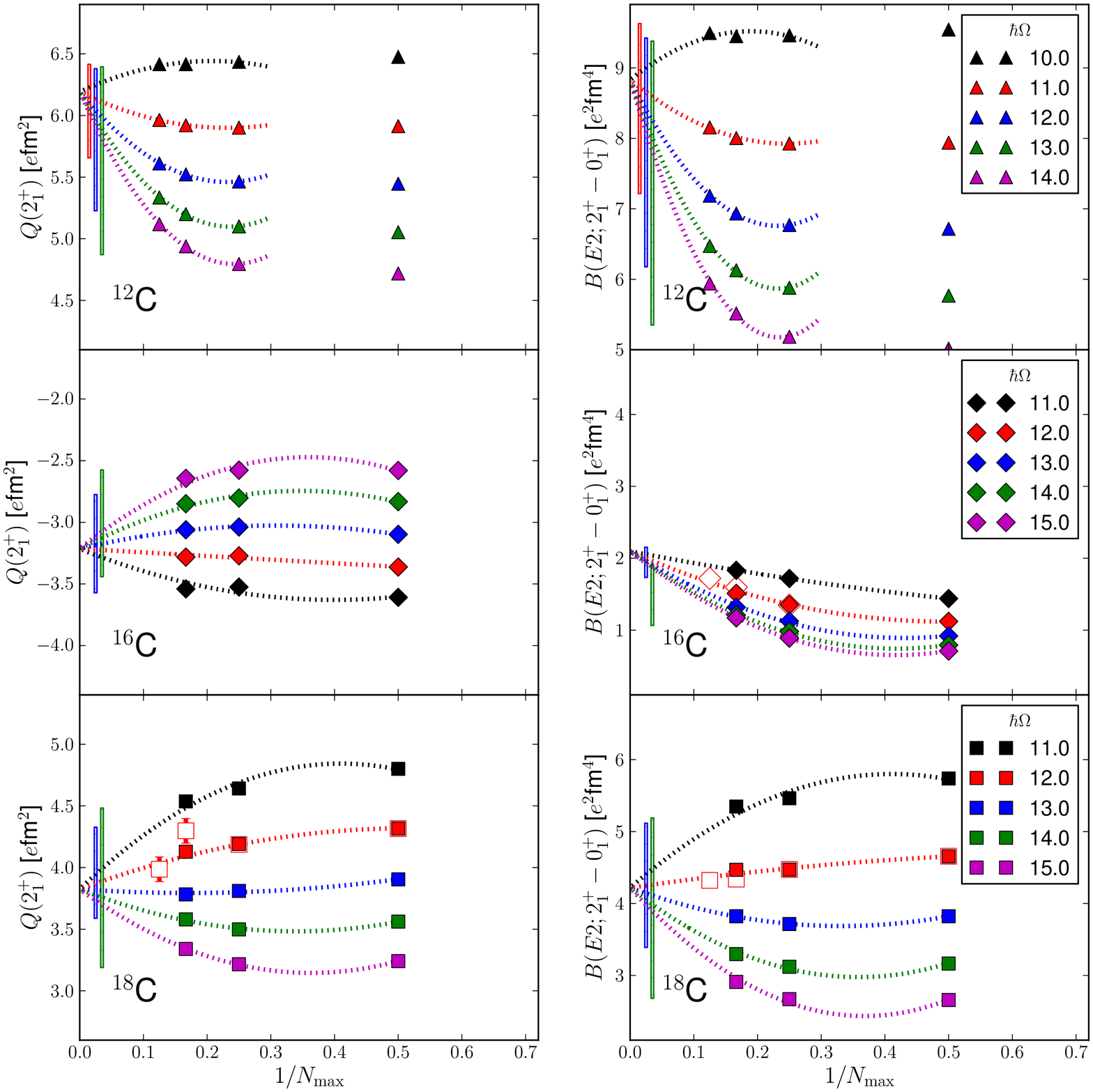}
  \caption{\co\ Model-space dependence of calculated E2 observables for
    \nuc{12,16,18}{C} 
    in the NCSM. Results obtained with
    the \cdb\ \nn\ potential are presented as a function of
    $1/\nm$. Filled (open) symbols correspond to full
    (importance-truncated) space results. Dotted lines
    correspond to constrained fits, and the bars correspond to linear
    fits as described in the text. See also
    Table~\ref{tab:qbe2}.%
    \label{fig:c_extrap}}
\end{figure}

Figure~\ref{fig:c_extrap} shows several examples of the extrapolation
procedures for quadrupole moments and $B$(E2) strengths. The data is the
same as in Figs.~\ref{fig:cAq},\ref{fig:cAbe2}. The results are plotted
as a function of $1/\nm$ for the selected range of HO frequencies. The
dashed lines correspond to the constrained fits to five
\ho-sequences. The bars correspond to the ranges from the linear
extrapolations. Starting from the right we have $\nm = (4,2) \,
(6,4), \, (8,6)$.  Numerical results for E2 observables for
all isotopes are presented in Table~\ref{tab:qbe2}. The range that is
presented from the linear fit corresponds to the largest \nm\ that was reached
for that particular isotope.
We note that the use of a range of frequencies usually include sequences that
converge from above and from below. This allows a more precise determination
of the extrapolated, final result. A particular exception to this
behavior is the $B$(E2) strength of \csi, for which all
sequences converge from below. This will be further commented below.
%
\subsection{\label{sec:twoplus}Systematics of electric quadrupole observables}
%
\begin{figure}[hbt]
  \centering
  \includegraphics*[width=8cm]
      {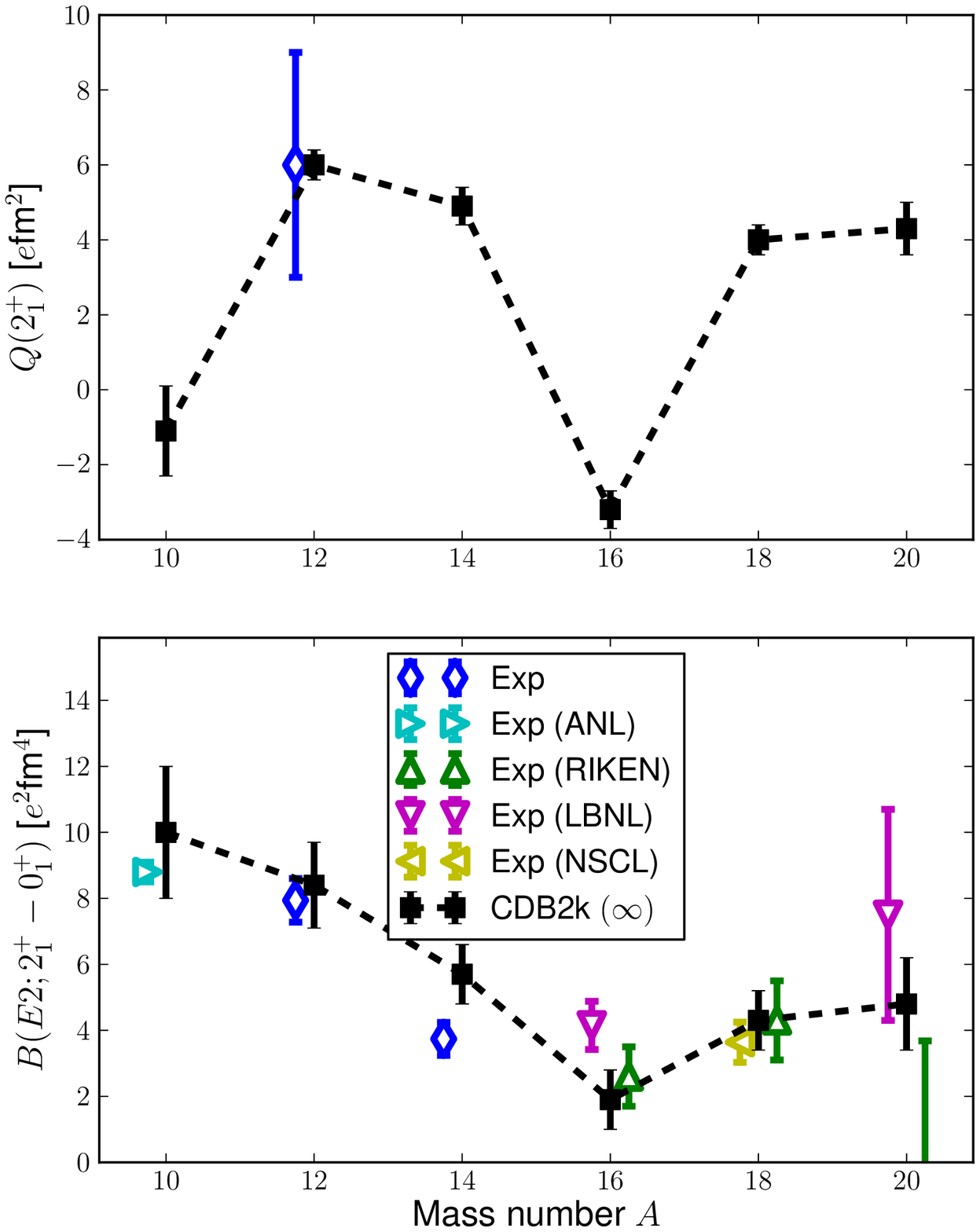}
\caption{\co\ NCSM calculated E2 observables of \nuc{10-20}{C} compared
  with experimental 
  results. See also Table~\ref{tab:qbe2}. NCSM results are obtained from
  linear  extrapolations in $1/nm$ (see text for details).%
  \label{fig:cA}}
\end{figure}
\begin{table*}[tbh]
  \caption{NCSM calculated E2 observables of \nuc{10-20}{C} compared
  with experimental results. NCSM results are obtained using the \cdb\
  interaction. The recommended values are obtained from two different
  extrapolation schemes: simple linear (lin.) and constrained fits
  (cons.) to sequences
  of \ho-frequencies (see text for details).
    \vspace*{1ex}%
    \label{tab:qbe2}}
  \begin{ruledtabular}
    \begin{tabular}{r|llc|c}
                 & \multicolumn{3}{c|}{$Q\left( 2^+_1 \right)$ [$e \mathrm{fm}^2$]}  \\
                 & \multicolumn{1}{c}{Th. (cons.~fit)} & \multicolumn{1}{c}{Th. (lin.~fit)} &\multicolumn{1}{c|}{Exp.} 
                 & Refs. \\
                 \hline
                 \cte    
                 & $ -1.1 \pm 1.2$ \ensuremath{^{a}} & --- & ---
                 &  \\
                 \ctw      
                 & $+6.2 \pm 0.2$ & $+6.0 \pm 0.4$ &  $+6 \pm 3$ 
                 & \cite{Vermeer:1983p23} \\
                 \cfo      
                 & $+4.7 \pm 0.4$ & $+4.9 \pm 0.5$ & --- 
                 & \\
                 \csi      
                 & $-3.2 \pm 0.3$ & $-3.2 \pm 0.5$ & --- 
                 & \\
                 \cei      
                 & $+3.8 \pm 0.2$ & $+4.0 \pm 0.4$ & --- 
                 & \\
                 \ctwe      
                 & $+4.3 \pm 0.6$ &  $+4.3 \pm 0.7$ &  ---
                 &  \\
                 \br
                 & \multicolumn{3}{c|}{\betz [$e^{2} \mathrm{fm}^{4}$]} \\
                 & \multicolumn{1}{c}{Th. (cons.~fit)} & \multicolumn{1}{c}{Th. (lin.~fit)} &\multicolumn{1}{c|}{Exp.} 
                 & Refs. \\
                 \hline
                 \cte    
                 & $10 \pm 2$\ensuremath{^{a}} & --- & $8.8 \pm 0.3$  
                 & \cite{McCutchan:2012in} \\
                 \ctw      
                 & $8.8 \pm 0.7$ & $8.4 \pm 1.3$ & $7.59 \pm 0.42$ 
                 & \cite{AjzenbergSelove:1990p1} \\
                 \cfo      
                 & $5.3 \pm 0.7$ & $5.7 \pm 0.9$ & $3.74 \pm 0.50$ 
                 & \cite{Crannell:1972p375} \\
                 \csi      
                 & $2.2 \pm 0.9$ & $1.9 \pm 0.9$ & $2.6 \pm 0.9, \, 4.15 \pm 0.73$ 
                 & \cite{Ong:2008p96, Wiedeking:2008p100} \\
                 \cei      
                 & $4.2 \pm 0.4$ & $4.3 \pm 0.9$ & $4.3 \pm 1.2, \,
                 3.64^{+0.55}_{-0.61}$ 
                 & \cite{Ong:2008p96, Voss:2012cm} \\
                 \ctwe      
                 & $4.8 \pm 1.1$ & $4.8 \pm 1.4$  & $<5.7, \,
                 7.5^{+3.2}_{-1.8}$
                 & \cite{Elekes:2009p82, Petri:2011p230} \\
   \end{tabular}
    \footnotespecial[1]{Strong mixing of the first two
                               \tp\ states. The estimate of \cte\ E2
                               observables is obtained by studying the
                               sums and ratios of results for both \tp\
                               states.}
  \end{ruledtabular}
\end{table*}
We focus now in particular on a discussion of the systematics of E2
observables in the chain of even carbon isotopes. First, we note that no
extrapolation was performed for the energy observables presented in
Fig.~\ref{fig:cAenergies}, although the magnitude of the \ho- and
\nm-dependence was indicated by the error bars. However, as discussed in
the previous section, a number of fit parameters are introduced in the extrapolation of
results for E2 observables to infinite model space.  In Fig.~\ref{fig:cA}
we compare the extrapolated theoretical results (linear fit) with the
experimental trends for the carbon chain of
isotopes. Numerical, extrapolated results, for both extrapolation
schemes, are presented in Table~\ref{tab:qbe2}.
It is obvious from Fig.~\ref{fig:cA} that our calculated \betz\ agrees rather
well with the most recent experimental data for the entire chain of
isotopes.

We note that the extrapolation of our \nuc{16}{C} \betz\ results is
particularly difficult. Unlike the trends for other carbon isotopes, the
$B$(E2) value increases with \nm\ in the whole investigated HO frequency
range (see Fig~\ref{fig:c_extrap}). This makes the upper bound less
constrained. 
As a consequence, for this case we introduce a systematic error to bring
the total uncertainty to $\pm 0.9$~$e^2$fm$^4$ in accord with the
neighbouring isotopes.  Our final recommended value is below the most recent
experimental results from LBNL~\cite{Wiedeking:2008p100,
  Petri:2012cw}. 
Furthermore, we note that our calculated
quadrupole moment for the first $2^+$ state of $^{16}$C is $Q
\approx-3.2$~$e$fm$^2$ while for $A=12,14,18,20$ we find the quadrupole
moment of the $2^+_1$ state to be positive. Concerning the quadrupole
moment of \cei\ we find that the threshold-extrapolation of this
observable in the IT-NCSM has a comparatively large uncertainty, which
shows up in the sizeable errorbars shown in the plot.

\begin{figure}[hbt]
  \centering
  \includegraphics*[width=10cm]
      {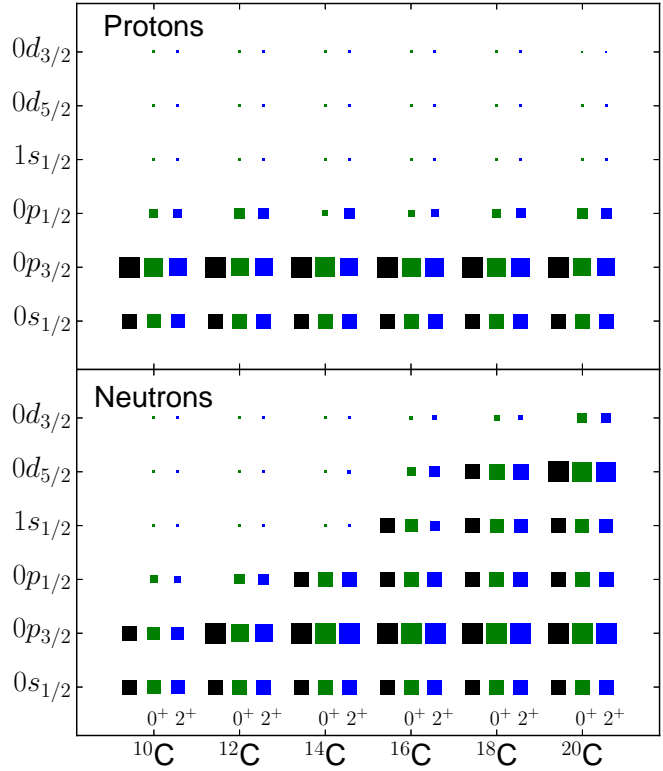}
\caption{\co\ Occupation numbers for the ground- (middle, green squares)
  and first \tp-state (right, blue squares) in
  \nuc{10-20}{C} obtained with the \cdb\ interaction. The area of the
  squares are proportional to the occupation numbers and can be compared
  with the unperturbed ground-state, shell-model
  occupation numbers (left, black squares).  We can note in particular the
  proton (neutron) excitation character of the \cfo(\csi) \tp\ state. %
  \label{fig:occs}}
\end{figure}

A qualitative understanding of these findings can be obtained by
studying the mean occupation numbers of different single-particle states in
the NCSM wave functions. In Fig.~\ref{fig:occs} these occupancies
are plotted for the ground- and first \tp-state for the whole range of
carbon isotopes. 
For comparison we show the ground-state occupation numbers that are
expected in an unperturbed shell model (non-interacting particles). We
show explicitly the occupation numbers up to the $sd$ shell. Not shown,
however, are occupation numbers for the $fp$ shell and beyond. They
extend up to 0.01--0.11 for both protons and neutrons.
The excitation mechanisms are quite obvious for \nuc{14,16}{C}. In \cfo\
the $2^+_1$ state corresponds to a proton excitation within the $p$
shell, while in \csi\ the $2^+_1$ state is obtained through a
re-configuration of neutrons in the $sd$ shell. The value of the
$B$(E2), for this particular case, will be quite sensitive to the fine
details of the re-configuration. Energy observables, on the other hand,
are not as sensitive to these small nuclear-structure details and can
therefore be expected to converge faster than the $B$(E2).

For \cte\ we observe a very strong mixing of the first two \tp\ states
using the \cdb\ interaction at small frequencies. To get at least crude
estimates of the E2 properties of the $2^+_1$ state we used a slightly
different extrapolation approach:
The ratios of, e.g., $Q(2^+_1)$ and $Q(2^+_2)$ was plotted for larger
frequencies where the mixing is not observed, while the sum was plotted
for the full range of frequencies. From such plots, for $Q$ and
$B(\mathrm{E}2)$ observables, we can deduce estimates for $Q(2^+_1)$ and
\betz\ and their uncertainties. These are included in
Table~\ref{tab:qbe2} and Fig.~\ref{fig:cA}.

We note that the different \nn\ interactions used in this study give
very similar isotopic trends for E2 observables, but with a consistently
smaller magnitude for the \inoy\ interaction. This observation is
connected to the anomalously large nuclear density generated by this
interaction found already in \nuc{4}{He}
calculations~\cite{Lazauskas:2004p198,Caurier:2006p197}.

Finally, a study of the characteristics of the second \tp\ state in
these isotopes strengthens the conclusion of the prominence of \csi\ in
the structural evolution of the chain of even carbon isotopes. The sign
of the quadrupole moment of this state, $Q(2^+_2)$, is reversed from
$Q(2^+_1)$. I.e., it's negative for all isotopes except for \csi\ (and
possible \cte). In
addition, as summarized in Table~\ref{tab:c_br}, the relative $B$(E2)
strength from this second \tp\ state to the ground state is much smaller
than that from the first \tp\ for all isotopes but \csi.
\begin{table}
  \caption{\label{tab:c_br} Relative $B$(E2) values for transitions
    among excited states of $^{14-20}$C. Results obtained in full
    \nm-space at fixed HO-frequency with the \cdb\ ($\ho=12$~MeV) and
    \inoy\ ($\ho=17-18$~MeV) \nn\ interactions are compared. $\nm =6$
    for \nuc{14-18}{C} and $\nm=4$ for \nuc{20}{C}.} 
\begin{ruledtabular}
\begin{tabular-cdddd}
& \multicolumn{2}{c|}{$\frac{B(E2;2^+_2 \rightarrow 0^+_1)}{\betz}$} 
& \multicolumn{2}{c}{$\frac{B(E2;2^+_2 \rightarrow 2^+_1)}{\betz}$} \\
$A$ & \multicolumn{1}{c}{\cdb}  &  \multicolumn{1}{c|}{\inoy}
& \multicolumn{1}{c}{\cdb}  &  \multicolumn{1}{c}{\inoy}\\
\hline
14 & 0.001 & 0.000 & 0.48 & 0.81 \\
16 & 2.2  & 0.30  & 2.0  & 0.79 \\
18 & 0.046 & 0.22\ensuremath{^{a}} & 0.029 & 1.7\ensuremath{^{a}} \\
20 & 0.017 & 0.035 & 0.12  & 0.28 \\
\end{tabular-cdddd} 
\footnotespecial[1]{For this particular interaction we observe
  considerable mixing between two $2^+$ states, with different
  structure, for certain choices of the HO frequency. These results are
  for $\ho=18$~MeV.}
\end{ruledtabular}
\end{table}
These findings are obtained with both \nn\ Hamiltonians used in this
study. However, the relative transitions from the second \tp\ in
\nuc{16,18}{C} stand out with clear differences in the predictions of
\cdb\ and \inoy, see Table~\ref{tab:c_br}. Note, however, that the
convergence of the second \tp\ state is computationally more
difficult, and therefore the statements on relative transition strengths
are based on runs performed at a single HO frequency. For \cei, in
particular, there is a strong \ho-dependence for the \inoy\ results that
makes the corresponding claim of a strong $2^+_2 \rightarrow 2^+_1$ E2
transition less robust. For \csi, however, the interaction dependence is
solid and intriguing. As the \inoy\ interaction often hints to possible
structural influence from \nnn\ forces we continue our study in the next
section with a more detailed investigation of the \csi\ structure using
Hamiltonians with realistic \nnn\ terms.
%
\subsection{\label{sec:spectrum}Higher-lying states of \nuc{16}{C} and
the role of the \nnn\ interaction}
\begin{figure}[tb]
  \centering
  \includegraphics*[width=10cm]
  {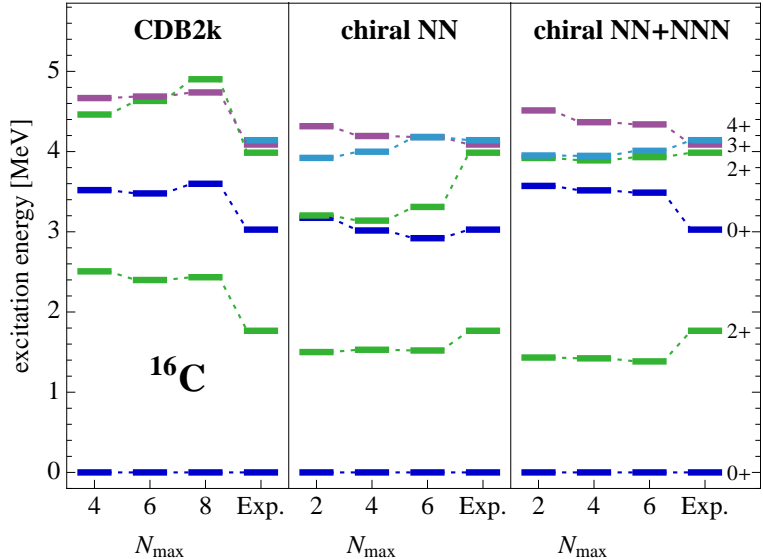}
  \caption{\co\ Excitation energies of the lowest states of
    $^{16}$C. Calculations using the Okubo-Lee-Suzuki-transformed \cdb\
    potential at $\ho=12$~MeV (left) and the
    SRG-evolved chiral \nn\ and \nn+\nnn\ interactions with 
    $\Lambda=1.88\,\mathrm{fm}^{-1}$ for $\hbar\Omega=16\,\mathrm{MeV}$
    (middle and right) are compared to experiment for different values
    of $N_{\max}$. The SRG-evolved chiral interactions include induced
    \nnn\ terms.
    \label{fig:c16_spectrum}}
\end{figure}

%
Transitions from higher excited states of \csi\ were studied in a recent
experiment~\cite{Petri:2012cw}. In particular, the transitions
$2_2^+ \rightarrow 2_1^+$, $4_1^+\rightarrow 2_1^+$ and $3_1^+
\rightarrow 2_1^+$ were observed. Interestingly, no transition from the
$2^+_2$ state to the ground state was seen. We performed additional
calculations with different Hamiltonians to study higher excited states
in $^{16}$C and their electromagnetic transitions. In
Fig.~\ref{fig:c16_spectrum}, we show the calculated and experimental
energy levels of $^{16}$C, and in Table~\ref{tab:c16_exctr} we summarize
our calculated $B$(E2) values among excited states normalized to
\betz. In particular, we compare results obtained with SRG-transformed
\nlonn\ and \nlonnn\ interactions (including the SRG-induced
three-nucleon terms in both cases as discussed in
Ref. \cite{Roth:2011p211}) calculated in the IT-NCSM, to those obtained with the two-body effective
\cdb\ interaction. A striking feature is a strong suppression of the
$2^+_2\rightarrow 0_1^+$ transition when the initial \nnn\ interaction
is included. The sensitivity to the presence of the \nnn\ interaction is
remarkable. The $2^+_2 \rightarrow 0_1^+$ transition is suppressed by a
factor of $\sim 7$ in the calculation with the \nnn\ compared to chiral
\nn\ only, and a factor of $\sim 20$ compared to \cdb. Clearly, the
calculation without the \nnn\ interaction contradicts the new MSU
experiment~\cite{Petri:2012cw} where indeed the $2^+_2
\rightarrow 0_1^+$ transition was not observed.
From Table~\ref{tab:c16_exctr} we observe that relative E2 transition
strengths obtained with the \nlonn\ interaction are similar to the ones
obtained with the \cdb\ interaction. Furthermore, we see from
Table~\ref{tab:c_br} that the relative $B$(E2) calculated with the
\inoy\ interaction (that mimics some \nnn\ effects) resemble results of
the \nlonnn\ Hamiltonian. 
The excitation energies of the five lowest $^{16}$C excited states are
also influenced by the \nnn\ interaction as seen in
Fig.~\ref{fig:c16_spectrum}. The agreement with the experimental spectrum is
quite reasonable in all presented cases, although slightly improved in
the calculation with the \nlonnn\ Hamiltonian.
\begin{table}
\caption{\label{tab:c16_exctr} Relative $B$(E2) values for transitions
  among excited states of $^{16}$C. Results obtained with the \cdb\ \nn\
  potential, the \nlonn, and the \nlonnn\ interaction are compared. For
    \cdb\ we use the Okubo-Lee-Suzuki effective interactions ($\hbar\Omega=12$ MeV,
    \nm=6) and for the chiral interactions we use SRG-evolved interactions
  ($\Lambda=1.88\,\mathrm{fm}^{-1}$, $\hbar\Omega=16\,\mathrm{MeV}$, \nm=6)
  including the induced three-nucleon terms.} 
\begin{ruledtabular}
\begin{tabular}{c c c c}
\br
$\frac{B(\mathrm{E2};J_i \rightarrow J_f)}{\betz}$ & \cdb\ & \nlonn\ & \nlonnn\ \\
\hline
$2^+_1\rightarrow 0^+_1$   & 1     & 1     & 1    \\
$2^+_2\rightarrow 0^+_1$   & 2.2   & 0.75  & 0.11 \\
$2^+_2\rightarrow 2^+_1$   & 2.0   & 1.7  & 0.65 \\
$3^+_1\rightarrow 2^+_1$   & 0.36  & 0.31  & 0.02 \\
$4^+_1\rightarrow 2^+_1$   & 0.89  & 0.69  & 0.80 \\
\br
\end{tabular} 
\end{ruledtabular}
\end{table}

From Table~\ref{tab:c16_exctr}, we also note a strong sensitivity of the
$3_1^+ \rightarrow 2^+_1$ transition to the presence of the \nnn\
interaction. The calculation with the \nlonnn\ Hamiltonian predicts a
strongly suppressed $B(\mathrm{E2;} \; 3_1^+ \rightarrow 2_1^+$)
transition. A transition between these states is observed,
however~\cite{Petri:2012cw}. Our calculation with the \nnn\
interaction predicts this transition to be of M1 character as seen from
Table~\ref{tab:bm1}. We also observe a sign change of the magnetic
moments of both the $2^+_1$ and the $2^+_2$ states in calculations with
the \nnn\ interaction included. The magnetic moment of the $3^+_1$ state
is unaffected, however. The sensitivity of the $2^+_1$ magnetic moment
to the \nnn\ interaction we also find in \ctwe\ (see
Table~\ref{tab:bm1}).

\begin{table}
  \caption{\label{tab:bm1} Magnetic dipole
    moments and $B$(M1) transition strengths
    of excited states in $^{16,20}$C. Results obtained with the \cdb\
    \nn\ potential and the SRG-evolved \nlonnn\ interaction are
    compared. $B$(M1) in $\mu_N^{2}$ and $\mu$ in $\mu_N$. Parameters as
    in Table~\protect\ref{tab:c16_exctr} with \nm=4 for \ctwe. The
    brackets indicate the uncertainties of the threshold extrapolation
    for the IT-NCSM.  }
\begin{ruledtabular}
\begin{tabular-cdrdr}
  & \multicolumn{2}{c|}{$^{16}$C}
  & \multicolumn{2}{c}{$^{20}$C} \\
  & \multicolumn{1}{c}{~} & \multicolumn{1}{c|}{chiral} 
  & \multicolumn{1}{c}{~} & \multicolumn{1}{c}{chiral}  \\   
  & \multicolumn{1}{c}{\cdb} & \multicolumn{1}{c|}{\nn+\nnn} 
  & \multicolumn{1}{c}{\cdb} & \multicolumn{1}{c}{\nn+\nnn}  \\
  \hline
  B($M1$; $2^+_2\rightarrow 2^+_1$) & 0.013 & 0.063 & 0.015 &   \\
  B($M1$; $3^+_1\rightarrow 2^+_1$) & 0.17 & 0.17 & 0.013 &   \\
  \hline
  $\mu(2^+_1)$ & 0.13 & -0.42 & 0.22 & 0.001(8) \\
  $\mu(2^+_2)$ & 1.3 & -0.79 & 0.58 & \\
  $\mu(3^+_1)$ & -3.2 & -3.1 & 0.016 & \\
\end{tabular-cdrdr} 
\end{ruledtabular} 
\end{table}

Overall, we find a strong sensitivity of the electromagnetic observables
in \csi\ to the details of nuclear Hamiltonian. Note, however, that we
don't employ two-body currents, and that these are expected to have a
non-negligible influence on magnetic dipole moments. Furthermore,
additional studies are needed regarding the effect of the similarity
transformation on this type of operator. But the long-range quadrupole
operator, that is the main target of this study, is not expected to be
much affected by this transformation. We conclude that more detailed
experimental study of higher excited states and their transitions will
be very useful.
%
\section{\label{sec:concl}Conclusion}
In summary, we have computed low-lying states of even carbon isotopes
with $A=10-20$ within the no-core shell model.  We have used several
accurate nucleon-nucleon (\nn) as well as \nn\ plus \nnn\ interactions
and calculated excitation energies of the lowest $2^+$ state, the
electromagnetic \betz\ transition strengths, the $2^+_1$ quadrupole
moments as well as selected electromagnetic transitions among higher
excited states. We use a truncated many-body model space, which however
can be systematically improved by 
increasing the cutoff. We employ two different similarity transformation
schemes to adapt the Hamiltonian to the available model space. 
The calculations do not include effective charges
or any other fitting parameters.  Note that the truncation of the
many-body basis used in the NCSM should in principle be followed by a
transformation of the transition operator that is consistent with the
renormalization of the Hamiltonian. Regarding long-range operators, such as
$Q$, this transformation is not expected to produce very different end
results for calculated
observables~\cite{Stetcu:2005p254,Anderson:2010p131}. In addition, the
small uncertainty associated with the approximation of using bare
operators is partly built into the error estimates that we obtain from
using several values of \ho\ and \nm.

We have presented full \nm-space results for energies and quadrupole
observables. In addition, we used two simple schemes to extrapolate the
quadrupole results to infinite model spaces. Additional fitting
parameters are introduced in these schemes that make extrapolated
results non-\emph{ab initio}.

Overall, we have found a consistent NCSM description of the \betz\
dependence on the mass number for the whole carbon isotopic chain from
$A=10$ to $20$. However, our extrapolated \betz\ values for \csi, with
different Hamiltonians, all underestimate the most recent experimental
measurements. A similar result was obtained by Ma~\emph{et
  al.}~\cite{Ma:2010p253} in a phenomenological approach. They used a
microscopic particle-vibration model to compute core polarization
effects. In their picture the reduced $B$(E2) strength in heavy carbon
isotopes can be traced back in particular to a strong quenching from
core polarization on $sd$-shell neutrons. In our approach, however,
there is no such separation into core and valence degrees of freedom.

In addition, we found a remarkable sensitivity of the
transition rates from higher excited states in \csi\ to the details of the
nuclear interactions. The \nlonnn\ interaction gives the excitation
spectrum of $^{16}$C in a slightly better agreement with experiment than
the \cdb\ \nn\ potential and, furthermore, the former interaction predicts
the suppression of the $2^+_2{\rightarrow} 0^+_1$ transition in
agreement with experimental observations. We found a strong
sensitivity of the magnetic moments of the $2^+_1$ state to the nuclear
interaction in $^{16}$C and $^{20}$C and even more so for the $2^+_2$
state in $^{16}$C.

The extrapolated NCSM results predict sign changes of the $2^+_1$ quadrupole
moments between different carbon isotopes. In particular, we predict a
negative quadrupole moment in $^{16}$C, a very small quadrupole moment
in $^{10}$C and a \betz\ value in $^{10}$C that is about the same as that
in $^{10}$Be. In $^{12}$C, we obtain $Q ( 2^+_1 ) = +6.0 \pm 0.4$~$e$fm$^2$. It
will be worth measuring these moments in the future. %
%
\begin{acknowledgments}
  We would like to thank A. Macchiavelli, P. Fallon. M. Wiedeking, and
  M. Petri for many useful discussions.  Support from the European
  Research Council under the FP7 (ERC grant agreement no.~240603), the
  Swedish Research Council (dnr.~2007-4078), the
  Deutsche Forschungsgemeinschaft through contract SFB 634, the
  Helmholtz International Center for FAIR (HIC for FAIR), and the BMBF
  (06DA9040I) is acknowledged. Support from the Natural Sciences and
  Engineering Research Council of Canada (NSERC) Grant No. 401945-2011
  is acknowledged. TRIUMF receives funding via a contribution through
  the National Research Council Canada.  Computing resources have been
  provided by the J\"ulich Supercomputing Centre and by LOEWE-CSC.
  Prepared in part by LLNL under Contract DE-AC52-07NA27344.
\end{acknowledgments}
\sectionreferences
\bibliography{forssen}
\end{document}